 \documentstyle[preprint,aps]{revtex}
\begin{document}
 \preprint{OITS-541,UM-P-94/47}
 \draft
\title{$\epsilon'/\epsilon$ And Anomalous Gauge Boson Couplings }
\author{Xiao-Gang He$^1$ and Bruce H.J. McKellar$^2$}
 \address{$^1$Institute of Theoretical Science\\
 University of Oregon\\
 Eugene, OR 97403-5203\\
 and\\
 $^2$Research Center for High Energy Physics\\
 School of Physics\\
 University of Melbourne \\
 Parkville, Vic. 3052 Australia}
\date{May, 1994}
\maketitle
\begin{abstract}
We study $\epsilon'/\epsilon$ in the Standard Model
and $\epsilon'/\epsilon$ due to anomalous $WW\gamma$
and $WWZ$ interactions 
as a function of the top quark mass.  In the Standard Model,
$\epsilon'/\epsilon$ is in the range $10^{-3} \sim 10^{-4}$
for the central value of top quark mass reported by CDF.
The anomalous gauge couplings can have large
contributions to the $CP$ violating $I=2$ amplitude in $K \rightarrow \pi\pi$.
Within the allowed regions for the anomalous gauge couplings,
$\epsilon'/\epsilon$ can be dramatically different from the standard model
prediction.
\end{abstract}
\newpage
\section{Introduction}
The $SU(2)_L\times U(1)_Y$ Standard Model (SM) of electroweak interactions
is in very good
agreement with present experimental data. The experimental data from LEP and
SLC and the
theoretical predictions in the SM for the gauge-fermion couplings agree  at the
1\% level or
better\cite{sm}. However, one of the most direct consequences of the SM, the
self-interaction of the gauge particles, the W, Z and photon,  characteristic
of nonabelian
gauge theories, has not been directly tested. It is important to study these
self-interactions to establish whether the weak bosons are gauge particles with
interactions  predicted by the SM, or gauge particles of  some extensions of
the SM which
predict different interactions at loop levels, or even non-gauge particles
whose
self-interactions at low energies are described by effective interactions.

Large uncertainties  
are introduced into studies of physics beyond the SM due to 
our lack of knowledge of the top quark mass $m_t$. D0 has put the lower bound
on $m_t$ to be 131 GeV\cite{d0t}.
CDF has announced evidence for the existence of top quark with a mass  of
$174\pm 10 ^{+13}_{-12}$ GeV\cite{cdft}.  If 
confirmed,
this information will allow us to make better predictions of new physics beyond
the SM. In this paper
we 
show how the information from CDF about the top quark mass helps the study of
the effect of anomalous gauge couplings on the CP violating parameter
$\epsilon'/\epsilon$ 
in comparison with the SM prediction.

In general there will be more gauge boson self-interaction terms than the tree
level
SM predicts.
The most general $WWV$
interactions with the W boson on shell, invariant under
$U(1)_{\mbox{\scriptsize{em}}}$, can be parametrized
as\cite{wwv}

\begin{eqnarray}
L_V &=& -ig_V[\kappa^V W^+_\mu W^-_\nu V^{\mu\nu} + {\lambda^V\over M_W^2}
W^+_{\sigma\rho}W^{-\rho\delta} V^{\;\sigma}_\delta\nonumber\\
&+&  \tilde \kappa^V W^+_\mu W^-_\nu \tilde V^{\mu\nu}
+ {\tilde\lambda^V \over M_W^2}W^+_{\sigma\rho}W^{-\rho\delta}\tilde
V^{\;\sigma}_\delta\nonumber\\
&+& g^V_1(W^{+\mu\nu}W^-_\mu - W^+_\mu W^{-\mu\nu})V_\nu
+g_4^VW^+_\mu W^-_\nu (\partial^\mu V^\nu + \partial^\nu V^\mu)\nonumber\\
&+& g_5^V \epsilon_{\mu\nu\alpha\beta}(W^{+\mu}\partial^\alpha W^{-\nu}
-\partial^\alpha W^{+\mu}W^{-\nu})V^\beta]\;,
\end{eqnarray}
where $W^{\pm\mu}$ are the W boson fields; $V$ can be the $\gamma$ or $Z$
fields;
$W_{\mu\nu}$ and $V_{\mu\nu}$ are
the $W$ and $V$ field strengths, respectively; and $\tilde V_{\mu\nu}
={1\over 2}\epsilon_{\mu\nu\alpha\beta}V^{\alpha\beta}$. The terms proportional
to $\kappa$,
$\lambda$, and $g^Z_{1,5}$ are CP conserving and $\tilde
\kappa$, $\tilde\lambda$ and $g_4^Z$ are CP violating.
For $V=\gamma$, $g_V = e$ and for $V=Z$, $g_V = g\cos\theta_W$.
$g_1^\gamma$ defines the W boson charge, one can always set it to  1.
In the SM at the tree level, $\kappa^V = 1$, $g_1^Z=1$, and all other couplings
in eq.(1) are zero. $\Delta \kappa^V = \kappa^V -1$, $\Delta g_1^Z = g_1^Z
-1$,
$\tilde \kappa^V$, $\tilde \lambda^V$, $g^V_4$ and $g^V_5$ are called
the anomalous gauge boson couplings.

There have been many experimental and theoretical studies of the anomalous
gauge boson couplings. Collider experiments at high energies have put
constraints on some of these couplings\cite{UA2,CDF}. It has been shown that
rare decays can provide important constraints\cite{chia,pet,he1,he2}. In
Refs.\cite{he1,he2}
using the recent data from CLEO on $b\rightarrow s \gamma$\cite{cleo} and data
on $K_L
\rightarrow \mu^+\mu^-$\cite{part}, constraints comparable or better than those
obtained in
collider physics were obtained. Rare $B$ decays may provide more stringent
constraints\cite{val}. The constraints from rare decays are better than those
obtained from the
$g-2$ of the muon\cite{mag}. In the literature the most  stringent constraints
on the anomalous
gauge boson couplings are from oblique corrections to the precision electroweak
experiments\cite{burg}.  The anomalous gauge coupling contributions to the
oblique corrections
are some times  quadraticly or even quarticly divegent. Care must be taken when
evaluating
these contributions.
Strictly, one should return to the underlying theories to remove the quartic
and quadratic  divergences\cite{burg}. For purely phenomenlogical studies,
we think the constraints from direct $W$ pair productions\cite{UA2,CDF}, and
rare
decays\cite{he1,he2} (the divergences here are at most logrithmic) are more
reliable.
 For the CP violating anomalous coupling, the best constraints
are from neutron and electron electric dipole moments\cite{de}.

In obtaining the bounds on the anomalous gauge boson couplings, most of the
analyses
assumed only one coupling is different from the SM tree level predictions.
A real underlying theory would produce more than just a single anomalous
coupling. If
the analyses were carried out including all anomalous couplings simultaniously,
the bounds would be much weaker. It is nevertheless interesting to find out if,
when
these stringent bounds are applied,  there are
 still large effects on other  processes. In this paper we will show that there
can be still
large effects on $\epsilon'/\epsilon$ from the anomalous $WW\gamma$  and $WWZ$
couplings.

The parameter $\epsilon'/\epsilon$ is a very important quantity
to study. It measures direct CP violation in $K\rightarrow \pi\pi$.
Experimental measurements are not conclusive at this stage\cite{epsilon},
\begin{eqnarray}
Re(\epsilon'/\epsilon) = \left \{ \begin{array}{ll}
(23\pm 6.5)\times 10^{-4}\;,\;\; &NA31\\
(7.4\pm6.0)\times 10^{-4}\;,\;\;& E731\\ \end{array} \right .
\end{eqnarray}
While the result of NA31 clearly indicates a non-zero $\epsilon'/\epsilon$, the
value of E731 is compatible with zero.  However, the two results are consistent
at the 2 standard deviation level.   The SM prediction for
$\epsilon'/\epsilon$ depends on the value of the top quark mass. It has been
shown that for a small top quark mass, the most important contributions to
$\epsilon'/\epsilon$ are from the strong penguin and isospin breaking due to
quark masses. For a large top quark mass, the electroweak penguins also
become important\cite{fr,buras}. In fact the sign of $\epsilon'/\epsilon$
may change for $m_t$ larger than 220 GeV.  If the top quark mass is indeed
about 174 GeV as reported by CDF, the electroweak penguin contribution will not
cancell the other
contributions completely. The predicted value for $\epsilon'/\epsilon$ is about
$10^{-3} \sim 10^{-4}$ which will
be within the reach of future experiment. We will then be able to find out
if there are other contributions to $\epsilon'/\epsilon$.
This illustrates the importance of knowing the $m_t$ in determining the physics
beyond the SM.

 The anomalous gauge intereactions are purely electroweak, so their
contributions to $\epsilon'/\epsilon$ will not affect the strong penguin but
may have significant effects on the eletroweak penguins. We will show that
the anomalous gauge couplings can change the result dramatically.

\section{Neutral flavor changing effective hamiltonian}

The effective Hamiltonian $H_{eff}$ for flavour
changing neutral currents with $\Delta F=1$, at the one loop level, is given by
\begin{eqnarray}
H_{eff} = H_{SM} + H_{AGC}\;,
\end{eqnarray}
where $H_{SM}$ is the SM contribution which can be find in Ref.\cite{lim}, and
$H_{AGC}$ contains the contributions from anomalous gauge couplings. It is give
by
\begin{eqnarray}
H_{AGC} &=& {G_F \over 2\sqrt{2}\pi}\sum_{i} V_{iq}V_{iq'}^*({e\over
8\pi}G(x_i)_A
\bar q' (m_{q'}(1-\gamma_5) + m_q(1+\gamma_5))\sigma_{\mu\nu} qF^{\mu\nu}
\nonumber\\
&+& \alpha_{\mbox{\scriptsize{em}}}Q_f H(x_i)_A \bar q' \gamma_\mu(1-\gamma_5)q
\bar f\gamma^\mu f
\nonumber\\
&+&\alpha_{\mbox{\scriptsize{em}}}\cot^2\theta_W F(x_i)_A \bar q' \gamma_\mu
(1-\gamma_5)q \bar f
\gamma^\mu (Q_f \sin^2\theta_W - T^3{1-\gamma_5\over 2}) f)\;,
\end{eqnarray}
with
\begin{eqnarray}
G(x)_A &=& - (\Delta \kappa +i\tilde \kappa) ({x\over (1-x)^2} +
{x^2(3-x)\over 2(1-x)^3}\ln x)\nonumber\\
&&-(\lambda +i\tilde\lambda) ({x(1+x)\over 2(1-x)^2} + {x^2\over (1-x)^3}\ln
x)\nonumber\\
H(x)_A&=&\Delta \kappa {x\over 4}\ln{\Lambda^2\over m_W^2} + \lambda
({x(1-3x)\over 2(1-x)^2} -
{x^3\over (1-x)^3}\ln x)\;,\\ F(x)_A &=& -\Delta g_1^Z {3\over 2}
x\ln{\Lambda^2\over m_W^2} +
g_5^Z ({3x\over 1-x} + {3x^2\over (1-x)^2} \ln x)\;.\nonumber
\end{eqnarray}
 For terms which are divergent in the
loop integral, we have just kept the leading terms. We used unitary gauge in
our calculations. Our first term in $H_A$ does not
agree with Ref.\cite{chia} where the author obtained a cut-off indenpendent
result. The term in
$H_A$ proportional to
$\Delta \kappa$ is similar to the term in the SM with $\kappa = 1$. In $R_\xi$
gauge, this term is gauge dependent\cite{lim}. In the unitary gauge this term
diverges.  This
gauge dependent term is cancelled by  terms from "box" and Z exchanges in
physical processes.
In our case because the coupling $\Delta
\kappa$ is anomalous, there are no terms coming from "box" and Z exchange to
cancel it.

The Hamiltonian in eq.(3) is the lowest nonvanishing order contribution to
flavor changing
neutral current. It has been show that QCD
corrections are important in the SM\cite{fr,buras,bruce}. QCD corrections
should be included in phenomenological analyses. To this end, we carry out the
leading log QCD
correction to the weak effective Hamiltonian.  The effective Hamiltonian at the
energy scale $\mu$
relevant to us can be written as
\begin{eqnarray}
H_{eff}^{\Delta S = 1} = {G_F\over \sqrt{2}}V_{ud}V^*_{us}
\sum_i C_i(\mu) Q_i(\mu)\;,
\end{eqnarray}
where i = 1,... 10 and
\begin{eqnarray}
C_i(\mu) = z_i(\mu) + \tau \tilde y_i(\mu)\;, \tau = -
V_{td}V_{ts}^*/V_{ud}V_{us}^*\;.
\end{eqnarray}
The coefficients $C_i$ satisfy the renormolization group equation to
the first order in $\alpha_s$ and $\alpha_{\mbox{\scriptsize{em}}}$,
\begin{eqnarray}
(\mu {\partial\over \partial \mu} + \beta(g) {\partial\over \partial g})
{\bf C}(\mu) = {1\over 2\pi}(\alpha_s(\mu)\gamma^{(s)T} +
\alpha_{\mbox{\scriptsize{em}}}(\mu)
\gamma^T){\bf C}(\mu)\;,
\end{eqnarray}
where  $\gamma^{(s)}$ and $\gamma$ are the anomalous dimension matrices which
were obtained in Ref.\cite{lus}.
The Wilson coefficients at the scale $\mu$ is obtained by first calculating
the coefficients at the scale $m_W$ and then using the renormalization group to
evolve down to the scale $\mu$.
In our calculation we will use experimental values for the CP
conserving amplitudes. We only need to calculate the Wilson coefficients
$\tilde y_i$ which are enter the calculation of CP violation. The four quark
operators are defined as
\begin{eqnarray}
Q_1 =&\bar s\gamma_\mu(1-\gamma_5)d \bar u\gamma^\mu (1-\gamma_5)u\;,
&Q_2 = \bar s\gamma_\mu(1-\gamma_5)u \bar u\gamma^\mu
(1-\gamma_5)d\;,\nonumber\\
Q_3 =&\bar s\gamma_\mu(1-\gamma_5)d \sum_q\bar q\gamma^\mu (1-\gamma_5)q\;,
&Q_4 = \sum_q\bar s\gamma_\mu(1-\gamma_5)q\bar q\gamma^\mu (1-\gamma_5)d\;,
\nonumber\\
Q_5 =&\bar s\gamma_\mu(1-\gamma_5)d \sum_q\bar q\gamma^\mu (1+\gamma_5)q\;,
&Q_6 = -2\bar s(1+\gamma_5)q \bar q (1-\gamma_5)d\;,\nonumber\\
Q_7 =&{3\over 2}\bar s\gamma_\mu(1-\gamma_5)d \sum_q Q_q\bar q\gamma^\mu
(1+\gamma_5)q\;,
&Q_8 = -3\sum_q Q_q \bar s(1+\gamma_5)q \bar q (1-\gamma_5)d\;,\nonumber\\
Q_9 =&{3\over 2}\bar s\gamma_\mu(1-\gamma_5)d \sum_q Q_q\bar q\gamma^\mu
(1-\gamma_5)q\;,
&Q_{10} = {3\over 2}\sum_q Q_q\bar s\gamma_\mu(1-\gamma_5)q\bar q\gamma^\mu
(1-\gamma_5)d\;,
\end{eqnarray}
Among these operators there are only seven linearly independent ones. We
 use $Q_{1,2,3,5,6,7,8}$ as the independent operators. The corresponding
coefficients
$y_{1,2,3,5,6,7,8}$ are given by
\begin{eqnarray}
y_1 &=\tilde y_1 - \tilde y_4 + {3\over 2} \tilde y_9 +
{1\over 2} \tilde y_{10}\;, &y_2 = \tilde y_2 +\tilde y_4 + \tilde y_{10}\;,
\nonumber\\
y_3 &= \tilde y_3 +\tilde y_4 -{1\over 2} \tilde y_9 -
{1\over 2}\tilde y_{10}\;, &y_i = \tilde y_i\;, i = 5,6,7,8\;.
\end{eqnarray}

The boundary conditions at $m_W$ for the Wilson coefficients in the SM can be
found in Ref.\cite{fr,buras,lim} which depend on the top quark mass. We will
not display them here. When the anomalous gauge coupling contributions are
included, due to the new contributions, the boundary conditions at $m_W$ for
the Wilson
coefficients are different from the SM. The new contributions will change
$\tilde y_{3,7,9}$. From eq.(4) we obtain the anomalous gauge boson coupling
contributions to
the Wilson coefficients at the $m_W$ scale,
\begin{eqnarray}
y_3(m_W) &=& - {\alpha_{\mbox{\scriptsize{em}}} \over 24\pi}
F_A(x_t)\;,\nonumber\\
y_7(m_W) &=& -{\alpha_{\mbox{\scriptsize{em}}}\over 6\pi}(H_A(x_t) +
\sin^2\theta_W F_A(x_t))\;,
\nonumber\\
y_8(m_W) &=& -{\alpha_{\mbox{\scriptsize{em}}}\over 6\pi} (H_A(x_t)
-\cos^2\theta_W F_A(x_t))\;.
\end{eqnarray}
The other coefficients are not changed from those of the SM. Note that the new
contributions to the effective Hamiltonian  depend only on
$\Delta \kappa^\gamma$,
$\lambda^\gamma$, $\Delta g_1^Z$ and $g_5^Z$. Contributions from the other
anomalous couplings are suppressed by factors like $O(m_{d,s}^2, m_K^2)/m_W^2$.
 We given in Table 1 and 2 the values for $y_i$ as a function of $m_t$
and the anomalous couplings.

\section{Contributions to $\epsilon'/\epsilon$}

The parameter $\epsilon'/\epsilon$ is a measure of CP violation in $K_{L,S}
\rightarrow 2 \pi$ decays. It is defined as
\begin{eqnarray}
{\epsilon'\over \epsilon} = i{e^{i(\delta_2-\delta_0)}\over
\sqrt{2}(i\xi_0 +\bar \epsilon)}\omega\left (
\xi_2 - \xi_0\right )\;,
\end{eqnarray}
where $\bar \epsilon\approx 2.26\times 10^{-3} e^{i\pi/4}$ is the CP violating
parameter in $K^0 - \bar K^0$,
$\delta_i$ are the strong rescattering phases,
$ \omega = |ReA_2/ReA_0| \approx 1/22$, and $\xi_i = \mbox{Im}A_i/ReA_i$. Here
$A_0$ and $A_2$
are the decay amplitudes with $ I = 0$ and $2$ in the final states,
respectively.

To separate different contributions to $\epsilon'/\epsilon$, we parametrize
$\epsilon'/\epsilon$ as
\begin{eqnarray}
{\epsilon'\over \epsilon} = \left ({\epsilon'\over \epsilon}\right )_6
(1 - \bar \Omega)\;,
\end{eqnarray}
where $(\epsilon'/\epsilon)_6$ is the contribution from $y_6$ which is
given by
\begin{eqnarray}
\left ({\epsilon'\over \epsilon}\right )_6 = {\omega\over 2\epsilon}
{G_F\over |A_0|}
y_6<Q_6> \mbox{Im}(V_{td}V^*_{ts})\;.
\end{eqnarray}
Here $<Q_i>_I$ is defined as $<Q_i>_I = <(\pi\pi)_I|Q_i|K>$.
The parameter $\bar \Omega$
contains several different contributions
\begin{eqnarray}
\bar \Omega = \Omega_{\eta+\eta'} + \Omega_{EWP} +\Omega_{octet} +\Omega_{27}
+\Omega_P\;
\end{eqnarray}
where $\Omega_{\eta+\eta'}$ is the contribution due to isospin breaking in the
quark masses which is estimated to be in the range $0.2\sim 0.4$\cite{lus,don}.
We will use  $\Omega_{\eta+\eta'}=0.25$ for illustration.
The other contributions
are defined as follows
\begin{eqnarray}
\Omega_{EWP} &=& {1-\sqrt{2}\omega\over \omega}  {y_7<Q_7>_2 +y_8<Q_8>_2
\over y_6<Q_6>_0} \;,\nonumber\\
\Omega_{octet}&=&  - {y_1<Q_1>_0 +y_2<Q_2>_0 \over y_6<Q_6>_0}\;,
\nonumber\\
\Omega_{27} &=& {1\over \omega} {(y_1+y_2)<Q_2>_2\over y_6<Q_6>_0}
\;,\nonumber\\
\Omega_P &=& - {y_3<Q_3>_0 +y_5<Q_5>_0\over y_6<Q_6>_0}\;.
\end{eqnarray}

The calculation of the hadronic matrix elements is the most difficut
task\cite{fr,buras,lus,hadron,wu}.  There is no satisfactory procedure for this
calculation at
 present. We will use the values in Ref.\cite{buras} in our tables and figures
for
illustration, and put our emphasis on
the effects of the anomalous couplings.  In Figure 1, we show the dependence of
$1-\bar \Omega$ as a function of $m_t$ and the anomalous gauge boson couplings.

\section{Discussion}

We show in Table 1 the SM predictions for the Wilson coefficients as a
function
of top quark mass $m_t$. In Table 2, we show the effects of
anomalous couplings on $y_{7,8}$ as functions of $m_t$ and the anomalous
couplings. It is clear
that the anomalous couplings can dramatically change the SM predictions.

In our numerical analyses of the effects of anomalous coupling on the  Wilson
coefficients,
we will assume that only one anomalous coupling  is non vanishing.
As have been mentioned before that this may not be true. We nevertheless carry
out the analysis
this way to
illustrate the effects of anomalous couplings on $\epsilon'/\epsilon$.
We use some values of the anomalous couplings
which are consistent with constraints from rare decays
because they are all derived from the effective Hamiltonian in eq.(3).
 The constraints from rare decays are top quark
mass $m_t$ dependent. Using
the recent CLEO bound on $b\rightarrow s \gamma$ at the 95\% CL\cite{cleo}, the
anomalous
couplings
$\Delta \kappa^\gamma$, $\lambda^\gamma$ are contrained to be in the range
$-2.2 \sim 0.35$ and $-6.7 \sim 1.1$ respectively for $m_t = 174$ GeV.
For larger $m_t$, the constraints are more stringent\cite{he1}.
These constraints are cut-off scale $\Lambda$ independent.
However the constrants from $K_L \rightarrow \mu^+\mu^-$ are cut-off dependent.
For $m_t = 174$ GeV the experimental data on
$K_L \rightarrow \mu^+\mu^-$ constrain $\Delta g_1^Z$ to be in the range
$-0.5\sim 0.1$ for cut-off scale $\Lambda = 1$ TeV. For larger $\Lambda$
the constraint is
more stringent\cite{he2}. $g^Z_5$ is constrained to be in the range $4 \sim
-1$.
the constraint on $g_5^Z$ is cut-off independent.  The specific values for the
anomalous
couplings are given in Table 2. We used values for the anomalous couplings
which are also
consistent with the constraints from collider phyiscs\cite{UA2,CDF}.

The anomalous couplings affect all the Wilson coefficients through
renormalization.
However, the effects on $y_{1,2,3,5,6}$ are less than 5\% and can be neglected.
The effects on $y_{7,8}$ are large. In Table 2, we show the effects of
anomalous couplings on $y_{7,8}$ as functions of $m_t$ and the anomalous
couplings.

 In Figure 1, we show the dependence of $1-\bar \Omega$ as a function
of $m_t$ and the anomalous gauge boson couplings. The anomalous gauge boson
couplings have a large effect on $\Omega_{EWP}$. The effect on other
contributions to $\Omega$ can be neglected.

Using the value $<Q_6>_0 = -0.255$ GeV$^3$ for $m_s = 0.175$ GeV, we have
\begin{eqnarray}
\left ( {\epsilon' \over\epsilon}\right )_6 \approx 8
\mbox{Im}(V_{td}V_{ts}^*)\;,
\end{eqnarray}
Here we have used $y_6 \approx -0.09$.
Using information from CP violation in $K-\bar K$ mixing and data from $B-\bar
B$
mixings\cite{part}, the allowed range for $\mbox{Im}(V_{td}V_{ts}^*)$ is
constrained to be in the
region $3\times 10^{-4} \sim  0.5\times 10^{-4}$ for $m_t$ varying from 100 GeV
 to 250 GeV. We see that $\epsilon'/\epsilon$ in the SM is
between $10^{-3}$ to $-3\times 10^{-4}$. There is a strong dependence on the
top quark mass $m_t$.
For the hadronic matrix elements used here, $\epsilon'/\epsilon$ changes sign
at about
$230$ GeV in the SM as mentioned before. If the top quark mass is determined,
the uncertainties for $\epsilon'/\epsilon$ will be greatly reduced.   The
physical top quark mass observed by experiments are different from the running
mass which we use in our calculation. A physical mass of 174 $GeV$
corresponding to a running mass about 165 GeV. For
$m_t = 165$ GeV, we find that $1-\bar\Omega = 0.3$ and $Im(V_{td}V_{ts}^*)$ is
in the range
$2\times 10^{-4}$ to $0.5\times 10^{-4}$. Therefore $\epsilon'/\epsilon$ is in
the range
$5\times 10^{-4} \sim 10^{-4}$ which will
soon be accessible to experiments at CERN and Fermilab.

There are, of course, uncertainties
due to our poor understanding of the hadronic matrix elements, erorr in
the QCD scale $\Lambda_4$ for four flavor effective quarks.
 In Ref.\cite{wu}, using a different set of hadronic matric elements, it is
found that the value for $1-\bar \Omega$ can vary a factor of two. It has
recently been shown that the next-to-leading order QCD corrections\cite{buras1}
can reduce the $\epsilon'/\epsilon$ about 10\% to 20\%. The uncertainty in
$\Lambda_4$ 
is $\pm 30\%$. In the above analysis, we have neglected contributions from
gluon dipole penguin operator of the form
$\bar q \sigma_{\mu\nu}\lambda^a(1-\gamma_5)q G^{\mu\nu}_a$. It has been shown
that to the leading order in chiral perturbation theory, this contribution
vanishes\cite{dhp}. Higher order chiral perturbation calculations indicate that
this contribution may enhance the value for $\epsilon'/\epsilon$ by about 10\%
for $m_t = 165$ GeV\cite{dhp,bert}.
When taking into account all the effects mentioned, we conclude that for
$m_t = 165$ GeV, $\epsilon'/\epsilon$ is in the range $10^{-3} \sim 10^{-4}$.

{}From Figure 1 it can be easily seen that the anomalous gauge couplings can
change the result dramatically. $\epsilon'/\epsilon$ can be much larger than
 the SM prediction
and  the value of $m_t$ where the sign change of $\epsilon'/\epsilon$
occurs can be significantly shifted. The change of sign for
$\epsilon'/\epsilon$
can occur for $m_t$ as small as 120 GeV for allowed values for the anomalous
gauge couplings. Future measurements on $\epsilon'/\epsilon$ will certainly
provide useful information about the anomalous gauge couplings.
  In Figure 1, we also show $1-\bar \Omega$ with the anomalous couplings set to
be $\pm 0.1$
for $\Delta \kappa^\gamma$ and
$\Delta g^Z_1$. We see that even with such small anomalous couplings, the
effects on
$\epsilon'/\epsilon$ are still sizeable. If we use the same bounds for
$\lambda^\gamma$ and
$g_5^Z$, the contributions are small (less  than 5\%).

We conclude that in the SM, $\epsilon'/\epsilon$ is predicted to be in the
range
$10^{-3}$ to $ 10^{-4}$ for $m_t$ = 165 GeV. The predicted values are within
the reach of future experiments. There can be large effects from the anomalous
gauge boson
interactions on $\epsilon'/\epsilon$, and hence measurement of
$\epsilon'/\epsilon$
can provide useful information about the anomalous gauge boson couplings.

 \acknowledgments
HXG was supported by the Department of Energy Grant No. DE-FG06-85ER40224.
BHJMcK was supported
in part by the Australian Research Council Grant number A69330949.
BHJMcK thanks Professor Deshpande and the Institute for Theoretical Science of
The University of Oregon for their
hospitality, which permitted this work to start.

\begin{table}
\caption{$y_i$ as a function of $m_t$ in the SM
at $\mu = 1$ GeV for $\Lambda_4 = 0.25$ GeV, $m_b = 5$ GeV, $m_c = 1.35$ GeV.}
\begin{tabular}{|c|r|r|r|r|r|}
$m_t$(GeV)  &140 &165 &180&200&240\\ \hline
$y_1$&0.041&0.039&0.038&0.037&0.033\\
$y_2$&-0.049&-0.049&-0.048&-0.048&-0.047\\
$y_3$&-0.020&-0.019&-0.019&-0.018&-0.017\\
$y_5$&0.012&0.012&0.012&0.013&0.013\\
$y_6$&-0.091&-0.092&-0.093&-0.093&-0.094\\
$y_7/\alpha_{em}$&-0.003&0.029&0.051&0.083&0.155\\
$y_8/\alpha_{em}$&0.081&0.121&0.149&0.188&0.278\\
\end{tabular}
\label{table1}
\end{table}

\begin{table}
\caption{$y_i$ as a function of $m_t$ and the anomalous gauge couplings
at $\mu = 1$ GeV for $\Lambda_4 = 0.25$ GeV, $m_b = 5$ GeV and $m_c = 1.35$
GeV, and the cut-off $\Lambda=1$ TeV.}
\begin{tabular}{|c|c|r|r|r|r|r|}
&$m_t$(GeV) &140 &165&180 &200&240\\ \hline
$\Delta \kappa= 0.2$&
$y_7/\alpha_{em}$&-0.036&-0.016&-0.003&0.0165&0.059\\
&$y_8/\alpha_{em}$&0.039&0.064&0.080&0.104&0.157\\
\hline
$\Delta \kappa= -0.5$&
$y_7/\alpha_{em}$&0.078&0.142&0.185&0.248&0.393\\
&$y_8/\alpha_{em}$&0.184&0.264&0.319&0.398&0.580\\
\hline
$\lambda = 1$&
$y_7/\alpha_{em}$&-0.034&-0.008&0.010&0.037&0.099\\
&$y_8/\alpha_{em}$&0.042&0.074&0.096&0.129&0.207\\
\hline
$\lambda = -3$&
$y_7/\alpha_{em}$&0.088&0.141&0.174&0.221&0.321\\
&$y_8/\alpha_{em}$&0.197&0.263&0.305&0.363&0.489\\
\hline
$\Delta g^Z_1= 0.05$&
$y_7/\alpha_{em}$&0.008&0.045&0.070&0.106&0.189\\
&$y_8/\alpha_{em}$&0.095&0.142&0.173&0.218&0.321\\
\hline
$\Delta g^Z_1= -0.5$&
$y_7/\alpha_{em}$&-0.121&-0.134&-0.143&-0.157&-0.190\\
&$y_8/\alpha_{em}$&-0.066&-0.082&-0.094&-0.111&-0.153\\
\hline
$ g^Z_5= 3$&
$y_7/\alpha_{em}$&-0.094&-0.079&-0.067&-0.048&-0.001\\
&$y_8/\alpha_{em}$&-0.033&-0.014&0.001&0.025&0.086\\
\hline
$\Delta g^Z_5= -0.5$&
$y_7/\alpha_{em}$&0.012&0.047&0.071&0.104&0.180\\
&$y_8/\alpha_{em}$&0.099&0.144&0.173&0.215&0.310\\
\end{tabular}
\label{table2}
\end{table}

\newpage
\begin{center}
Figure Captions
\end{center}

Figure 1. $1-\bar \Omega$ as a function of $m_t$ and anomalous gauge boson
couplings. Different values for $\Delta \kappa^\gamma$, $\lambda^\gamma$,
$\Delta g_1^Z$ and $g_5^Z$ are
used in Figures a,b,c, and d, respectively. In each of the figures all other
anomalous couplings are set to be zero.

\end{document}